# Photocurrents in a Single InAs Nanowire/ Silicon Heterojunction

*Andreas Brenneis[1,2], Jan Overbeck[1,2], Julian Treu[1], Simon Hertenberger[1], Stefanie Morkötter[1], Markus Döblinger[3], Jonathan J. Finley[1,2], Gerhard Abstreiter[1,4], Gregor Koblmüller[1], and Alexander W. Holleitner[1,2]*

1) Walter Schottky Institut and Physik-Department, Technische Universität München, Am Coulombwall 4a, 85748 Garching, Germany.
2) Nanosystems Initiative Munich (NIM), Schellingstr. 4, 80799 Munich, Germany.
3) Department of Chemistry, Ludwig Maximilians Universität München, Butenandtstr. 11, 81377 Munich, Germany.
4) Institut for Advanced Study, Technische Universität München, Lichtenbergstrasse 2a, 85748 Garching, Germany

## Abstract

We investigate the optoelectronic properties of single indium arsenide nanowires, which are grown vertically on p-doped silicon substrates. We apply a scanning photocurrent microscopy to study the optoelectronic properties of the single heterojunctions. The measured photocurrent characteristics are consistent with an excess charge carrier transport through mid-gap trap states, which form at the Si/InAs heterojunctions. Namely, the trap states add an additional transport path across a heterojunction, and the charge of the defects changes the band bending at the junction. The bending gives rise to a photovoltaic effect at a small bias voltage. In addition, we observe a photoconductance effect within the InAs nanowires at large biases.

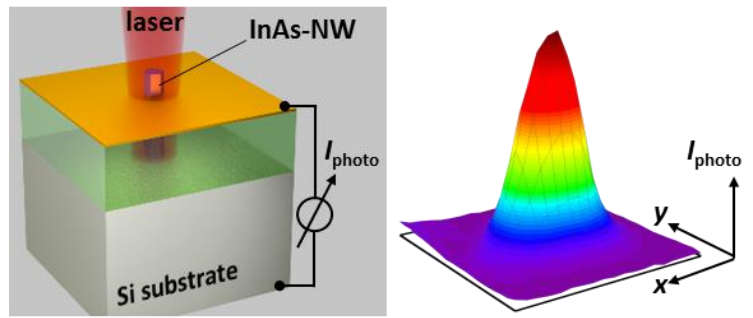

The integration of III/V semiconductors on a silicon platform offers great advantages for future electronic and optoelectronic devices. Recently, the epitaxial growth of vertically aligned III/V nanowires (NWs) on silicon substrates has been established.[1–9] Such nanowire-based circuits have been demonstrated for photovoltaic applications,[10–18] photodetectors,[19,20] transistors,[21,22,22] tunnel diode junctions,[23–26] and lasers.[27–29] Amongst the III/V family, the small band gap of InAs of 0.36 eV at room temperature and its high electron mobility render InAs as an interesting candidate for electronic and optoelectronic devices. For all applications, an extensive understanding of the heterojunction between InAs and Si is essential.[30,31] A simple approximation of the band alignment at the heterojunction can be derived from the material specific, bulk parameters like electron affinity, bandgap, and doping. Consequently, the formation of a space charge layer depends on the difference of the work functions of the two materials. However, defects are likely to exist at the interface of two different materials. We demonstrate that defects need to be considered to describe the optoelectronic properties of Si/InAs heterojunctions. In particular, charged defects and interface dipoles can modify the potential landscape at the junction, and they can add a new transport path for charge carriers across the heterojunction.

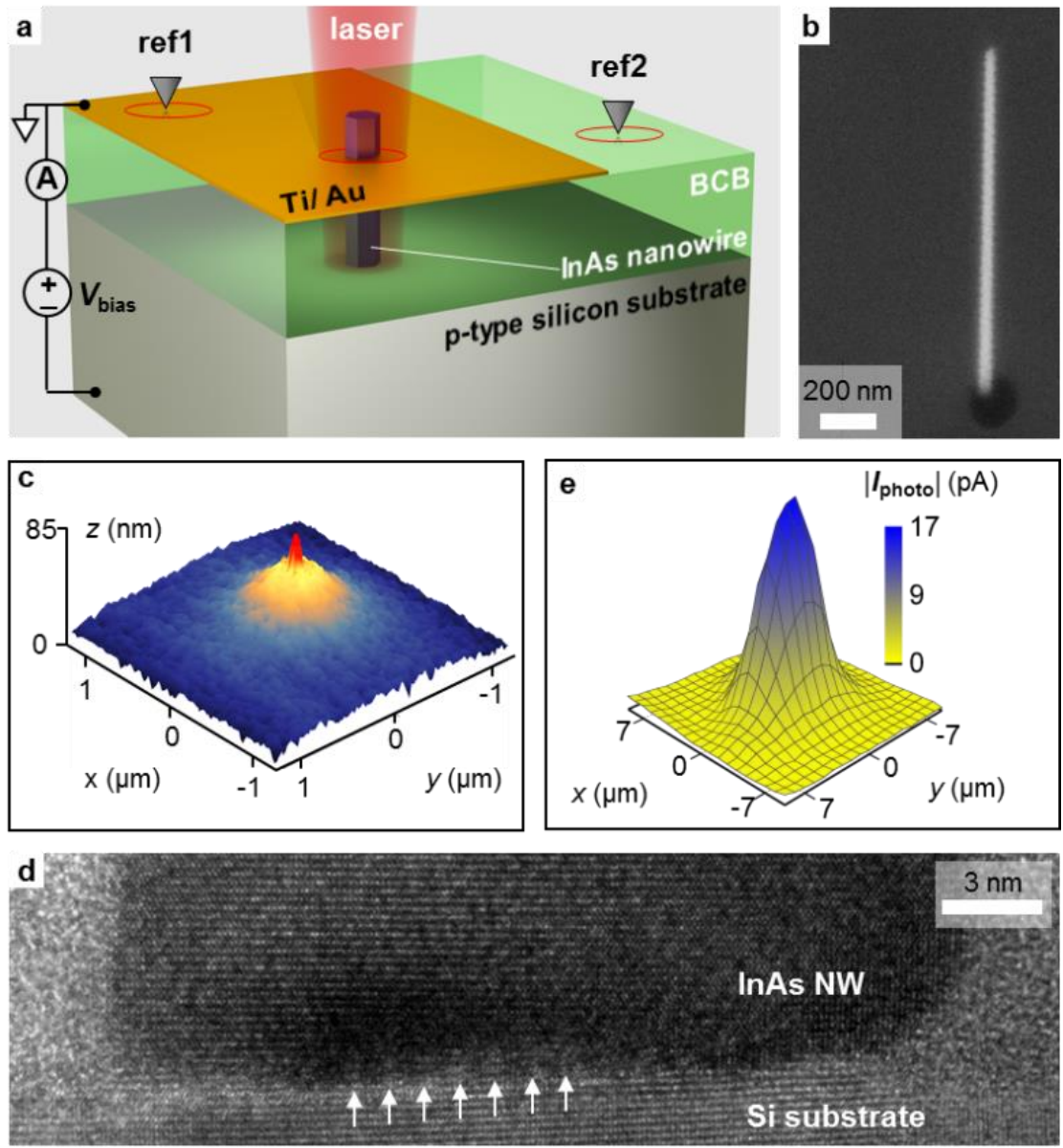

Figure 1 **a.** Sample layout. a focused laser excites a single InAs nanowire (NW) grown on a p-type Si (111) substrate. The NW is embedded in an insulating polymer (Bisbenzocyclobutene, BCB) with a semitransperant top contact. The photocurrent is measured from the top contact to the Si substrate. Reference excitation positions "ref1" on the top contact and on the polymer "ref2" are schematically indicated. **b.** Scanning eletron microscope (SEM) image of a molecular beam epitaxy grown InAs-NW on a Si (111) substrate. The Si substrate is covered with silicon oxide. The NW nucleates from a prepatterned hole in the silicon oxide. **c.** Atomic force microscope (AFM) image of a NW tip that is exposed from the BCB polymere before metalization. **d.** Transmission electron microscope (TEM) images of the heterojunction betwenn InAs NWs and the Si substrate. The images are recorded in slightly tilted view to visualize the strain patterns (white arrows). **e.** Photocurrent map of a single contacted InAs-NW.

We investigate single InAs nanowires grown *via* molecular beam epitaxy on p-type silicon substrates at the telecommunication wavelength 1470 nm. We use scanning photocurrent microscopy to locally measure the optically induced currents (Figure 1a). Our results are consistent with the transport of photogenerated charge carriers across dislocation-mediated trap states that form at the Si/InAs heterojunction. This scenario is very different from an ideal defect-free heterojunction. Based on our optoelectronic findings, we elaborate insights into the band alignment at the Si/InAs heterojunction. This alignment is equally essential for transport devices like transistors, optoelectronic devices like photodetectors, and solar cells.[13,14,22,25,30–34] All of them rely on the excess charge carrier transport across the Si/InAs heterojunction. We further discuss the influence of the defects in the silicon substrate on the near infrared photoresponse. Our results may prove essential to develop and build future solar cells and optoelectronics based on single III/V-nanowires which are integrated on silicon platforms.

## Results and Discussion

### Sample Fabrication

Figure 1a illustrates the device geometry used to measure the spatially resolved photocurrent of single InAs nanowires (NWs). Unintentionally n-type doped InAs NWs are grown by molecular beam epitaxy (MBE) on a p-type Si (111) substrate (1-30 Ωcm), where a thin lithographically prepatterned SiO$_2$ mask template induces site-selective growth (Figure 1b and methods).[4] An insulating polymer (Bisbenzocyclobuten, BCB) embeds the NW and it further supports a semitransparent metal top contact (Figure 1a and methods). Atomic force microscope (AFM) images taken prior to the metallization reveal that the InAs NWs stick through the BCB spacing layer (red tip in Figure 1c).

Figure 1d shows a cross-sectional high-resolution transmission electron microscope (TEM) image of the heterojunction of an InAs NW and the Si substrate, which was grown under similar conditions. The image highlights crystalline InAs and Si materials with an atomically abrupt interface. Due to the high lattice mismatch of 11.6 % between InAs and Si, periodic misfit dislocations exist at the interface which are indicated as periodic brightness variations (indicated by white arrows).[22,25,30,32,34] Such dislocations are known to cause trap states in the bandgap of InAs [30] and they are important to describe the charge carrier transport across the heterojunction.[30,32,33] We note that the InAs NWs crystallize in wurtzite (WZ) phase with a high density of stacking defects, which is commonly observed under the selected catalyst-free MBE growth conditions.[9,35] As we will discuss below, the charge trap states at the Si/InAs heterojunction can dominate the overall optoelectronic response of the devices.

### Scanning photocurrent microscopy

We apply a bias voltage $V_{\text{bias}}$ to the substrate, while the potential of the top contact is defined as ground (Figure 1a). The sample is excited from the top with a focused laser beam with a wavelength of $\lambda$ = 1470 nm (0.84 eV). The laser irradiation is modulated with a chopper wheel in the kHz-frequency range, and a lock-in amplifier records the optically induced current $I_{\text{photo}}$ at the trigger frequency. The position of the laser spot is scanned across the sample with a piezo stage to record two-dimensional photocurrent maps (methods). Figure 1e shows such a map at the position of a single contacted InAs-NW. As can be seen, the photocurrent amplitude is strongly enhanced at the position of the NW. The underlying optoelectronic processes are discussed below.

### Dark currents

Without laser illumination, we observe a diode-like behavior for the contacted, single InAs-NWs in the temperature range of 250 K < $T$ < 300 K (Figure 2a). The temperature-dependent reduction in forward direction ($V_{\text{bias}}$ > 0 V) can be explained by a thermionic emission model. In particular, only holes of the p-doped Si, that have sufficient energy to overcome the thermionic barrier $\phi_{\text{B}}$ (see band alignment in Figure 2b), contribute to the current density $j_{\text{TE}}$. In contrast, the thermionic emission of electrons from the InAs can be neglected because of the high conduction band offset $\Delta E_{\text{C}}$. The thermionic emission current density is given by [36]

$$j_{\text{TE}} = A^* T^2 \exp\left(-\frac{\phi_{\text{B}}}{k_B T}\right) \cdot \left[\exp\left(\frac{eV_{\text{bias}}}{k_B T}\right) - 1\right], \quad (1)$$

with $A^* = \frac{4\pi e m^* k_B^2}{h^3}$ the Richardson constant, $e$ the electron charge, $m^*$ the effective hole mass, $k_B$ the Boltzmann constant, $h$ the Planck constant, and $T$ the bath temperature. In order to fit the current voltage characteristic in Figure 2a, we assume a model circuit as sketched in the inset of Figure 2a. We attribute the shunt current $V_{bias}/R$ to thermally excited minority charge carriers (Figure in supporting material). Equation 1 can be transformed to

$$\ln\left(\frac{j_{TE}}{T^2}\right) \propto -(\phi_B - V_{bias}) \cdot \frac{e}{k_B T}. \quad (2)$$

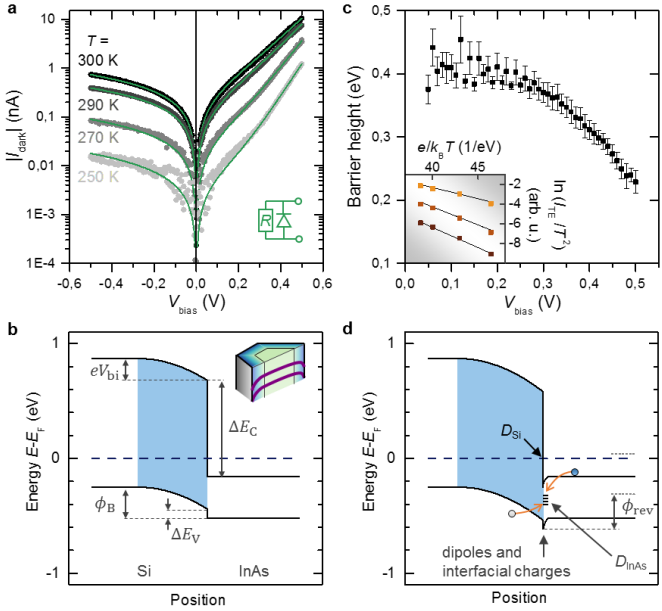

Figure 2 **Current voltage characteristics without illumination and band alignment a.** Current voltage characteristics of a single contacted InAs NW for 250 K < $T$ < 300 K. **b.** Band alignment of the bulk p-Si/n-InAs heterojunction based on bulk material constants. Inset: Band bending (for both CB and VB) across radial profile of the InAs-NW. The surface pinning of InAs leads to an accumulation of electrons at the surface; holes tend to move towards the core. **c.** Thermionic barrier height $\phi_B$, extracted from the temperature behavior of the forward current for 250 K < $T$ < 300 K. The inset exemplarily shows the Arrhenius plots of $I_{TE}/T^2$ for $V_{bias}$ = 0.1, 0.3, and 0.5 V (top to bottom) **d.** Schematic band structure accounting for dipoles and interfacial charges at the heterojunction. Interfacial charges lead to an opposite band bending in the InAs. Under forward bias, electrons from the InAs can recombine with holes from the Si via the trap states $D_{InAs}$ in the InAs at the heterojunction (orange arrows). The Fermi level pinning at the surface states of the Si is indicated with an arrow labeled $D_{Si}$.

Thus, the barrier height $-(\phi_B - V_{bias})$ can be extracted from the slope in an Arrhenius plot of $I_{TE}/T^2$ (inset of Figure 2c), where $I_{TE}$ is the corrected thermionic current contribution $I_{TE} = I_{dark} - V_{bias}/R$, where the shunt current was subtracted. The thermionic barrier $\phi_B$ can then be interpolated from the data for $V_{bias}$ = 0 V, as shown in Figure 2c. We find a barrier height $\phi_B$ of 0.4 - 0.5 eV, which is slighter larger than the value 275 meV found by Björk et al..[32] However, the lower bound of our value is given by the data for $V_{bias}$ < 0.25 V. Here, the error is relatively large because of the lower signal to noise ratio for small currents (especially at low temperatures). The upper bound of the barrier height is extracted for $V_{bias}$ > 0.25 V, where the barrier height becomes comparable to the applied voltage. In turn, the voltage that drops in the Si substrate cannot be neglected. Thus, the simple thermionic model can no longer be applied for higher bias voltages. From this, we deduce a thermionic barrier height of $\phi_B \approx 0.4$ eV for the investigated Si/InAs heterojunctions.

**Band alignment**

We estimate the band profile of the Si/InAs heterojunction using bulk parameters for zinc-blende InAs and Si (Figure 2b). This estimation is adequate for the discussion of our photocurrent results for two reasons. (i) The lattice parameter difference between the Si substrate and InAs is very similar for wurtzite (WZ) and zinc-blende (ZB) InAs.[37] (ii) The bandgap of WZ InAs is only ≈ 36-50 meV larger than compared to the bandgap of the ZB phase [38–41] and this bandgap difference is negligible compared to the conduction band offset between InAs and Si. The valence band edge of WZ InAs is expected to be shifted 39- 46 meV up with respect to ZB InAs in a direct WZ/ ZB InAs interface.[38,39,41] Such a higher value of the valence band edge for the WZ case is in agreement with all of our argumentation and conclusions. The assumed material parameters are: Electron affinity $\chi_{Si}$ = 4.05 eV, $\chi_{InAs}$ = 4.90 eV, bandgap $E_{g,Si}$ = 1.12 eV,[36] $E_{g,InAs}$ = 0.36 eV,[36] Fermi level $E_F - E_{vb,Si}$ = 0.25 eV.[42] We can further assume the position of the Fermi level $E_F$ at the surface of the InAs-NWs to be located above the conduction band $E_{cb,InAs}$ of the NWs (inset of Figure 2b). For instance, $E_F$ has been reported to be pinned 0.13 - 0.20 eV above $E_{cb,InAs}$ at the surface.[43–45] The unintentional doping concentration of the InAs-NWs used in this paper is ≈ $1 \cdot 10^{17}$ cm$^{-3}$.[46] Furthermore, the work function of the p-type Si substrate is larger than the work function of InAs (c.f. caption of Figure 2). Therefore, there will be a depletion region in the space charge region of the Si substrate (shaded area in Figure 2b). In this ideal model, the dimension of the space charge region in InAs is negligible because of the much higher charge carrier concentration in the InAs nanowire.

Typically the interface of Si/InAs heterojunctions is not ideal. Periodic misfit dislocations can be present close to the heterojunction due to the large differences in lattice constants between InAs and Si (Figures 1d).[4] These dislocations can cause charged trap states (acceptor states) within the bandgap of InAs,[30] which tend to accumulate electrons on both sides of the Si/InAs heterojunction. As a result, the slope of the band bending can change its sign close to the heterojunction (Figure 2d).[47–49] For clarity, we will use the abbreviation $D_{InAs}$ for the mid-gap trap states in InAs. It is anticipated that dipoles at the heterojunction change the band offsets $\Delta E_V$ and $\Delta E_C$. The influence of all these effects is qualitatively summarized in Figure 2d. In particular, the interfacial trap states add an additional transport path for electrons from the conduction band of the InAs-NW to recombine with holes in the valence band of the Si (orange arrows in Figure 2d).

In addition to the misfit-induced defects $D_{InAs}$ in the InAs-NW, mid-gap defects in the Si can arise at the surface either from a native silicon oxide layer, broken Si-H bonds, or HF etching.[36,50] HF etching is used to define the nucleation position of the NWs in the silicon oxide mask (methods) and to passivate the entire silicon surface prior to the encapsulation in BCB. Therefore, we suggest that the HF-induced defects are the primary candidates for mid-gap defect states in the Si. We introduce the abbreviation $D_{Si}$ for the mid-gap defects in the Si substrate. The Fermi level at the surface of Si can be pinned by these mid-gap defects $D_{Si}$ as indicated in Figure 2d.

## Optoelectronic results

In the following, we present the photocurrent characteristics measured with the laser focused onto a single InAs NW. We further compare these results with excitation positions "ref1" and "ref2", that are both ≈ 30 μm away from the NW (Figure 1a). The position at the NW and at "ref1" exhibit a semitransparent top contact, which reduces the laser irradiation by a factor of ≈ 25. Therefore, the laser power at the reference position "ref2" is expected to be 25 times higher than for the positions at the NW and at "ref1". Despite of this, the amplitude of the photocurrent is enhanced for an excitation at the NW position for most of the applied bias voltages (Figure 3a). Additionally, we observe a negative open circuit voltage $V_{oc,NW}$ ≈ -21 mV for the excitation position at the NW (Figure 3a). For the positions "ref1" and "ref2", we observe a positive value of $V_{oc}$. Since we excite with a laser energy below the bandgap energy of silicon, the photocurrent contribution for the position on the silicon ("ref1" and "ref2") can be understood from an excitation of the defect states $D_{Si}$ at the Si surface.[51] Such mid-gap defect states enable an optical excitation from the valence band or to the conduction band.[50] The density of these states can be reduced by a hydrogen termination of the dangling bonds,[52] as it is done in this work by a HF dip etch prior to the BCB spin coating. However, even if the time between surface termination and encapsulation of the surface is kept minimal (few minutes), we measure a small photocurrent signal at the positions "ref1" and "ref2". Additionally, as mentioned above, HF etching can cause defects in Si as well.[50] For $|V_{bias}|$ > 0.1 V, the photocurrent contribution at the position of the NW is larger than the contribution from the silicon "ref1" (Figure 3a). We interpret this finding in a way that the photocurrent at the NW position is dominated by the absorption within the InAs NW and not within the silicon substrate, as will be discussed below.

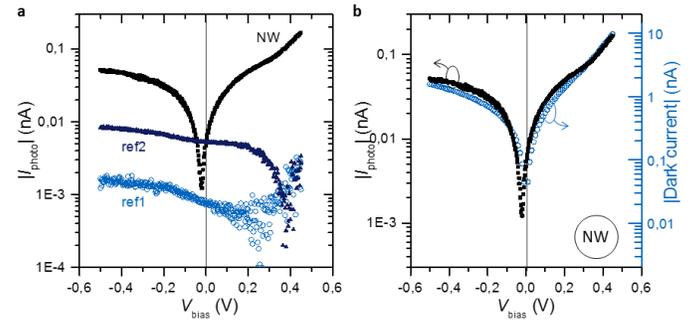

Figure 3 **Photocurrent: a.** Photocurrent *vs.* bias voltage for the excitations positions NW, "ref1", and "ref2" as defined in Figure 1a. Experimental parameters are $P_{laser}$ = 1.87 mW, $f_{chopper}$ = 833 Hz, $T$ = 298 K. **b.** Photocurrent at the position of the NW as in figure a. superimposed with the dark current at the same temperature $T$ = 298 K.

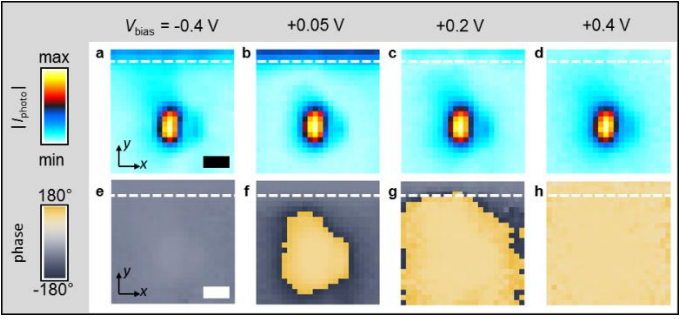

Figure 4 **Photocurrent maps of a single InAs-NW: a.-d.** show the normalized amplitude of the photocurrent and **e.-h.** the respective phase. The dashed lines indicate the boundary of the semitransparent top contact. Scale bar 5 µm. Experimental parameters are $P_{laser} \approx 1.9$ mW, $f_{chopper} \approx 833$ Hz, and $T = 298$ K.

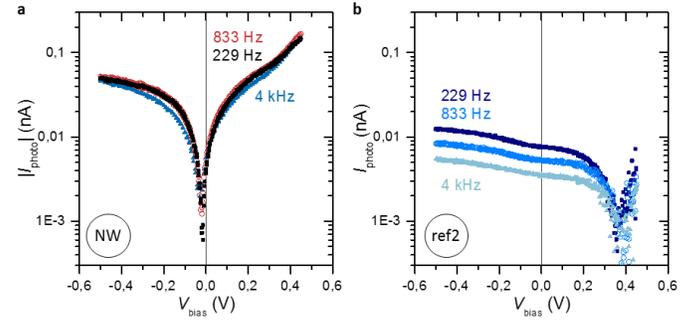

Figure 5 **Chopper frequency dependence of the photocurrent: a.** Chopper frequency dependence for the excitation position on the InAs-NW. **b.** Same dependence on the reference position "ref2". Experimental parameters are $P_{laser} = 1.9$ mW and $T = 298$ K.

Based on the ideal bulk band alignment depicted in Figure 2b, one expects a positive open circuit voltage for the excitation of charge carriers in the silicon. In this model, electron hole pairs, which are photogenerated in the Si, will be separated in the space charge region (shaded area in Figure 2b). Thus electrons and holes will drain through the NW and Si substrate, respectively. The distance of all excitation positions to the space charge region at the heterojunction is much smaller than the diffusion length of excess charge carriers in Si, which is in the order of 1 mm for the used Si-substrate.[36] Indeed, for the position "ref1", the open circuit voltage is $V_{oc,ref1} > 0.2$ V. To increase the signal to noise ratio, we choose the position "ref2" without the top contact and therefore, with a larger laser irradiation. Indeed, we find a positive open circuit voltage of $V_{oc,ref2} \approx +0.38$ V with a clear signal to noise ratio (Figure 3a). The value is on the order of the thermionic emission barrier $\phi_B \approx 0.4$ V (Figure 2c), which corresponds to the flat band condition of the Si/InAs heterojunction. In particular, we interpret the difference $\phi_B - V_{oc,ref2}$ to represent the offset $\Delta E_V$ between the valence bands of the Si/InAs heterojunction. In contradiction to the ideal model of Figure 2b, we observe a negative open circuit voltage when the laser is positioned at the NW (Figure 3a). From this observation, we can conclude that the optoelectronically dominant band bending at the Si/InAs heterojunction must be opposite to the band bending within the space charge region of the silicon. This opposite band bending is consistent with the band alignment depicted in Figure 2d, which considers the defect and trap states at the heterojunction. We note that the photocurrent characteristic measured at the position of the NW resembles the dark current of the overall device despite of the offset voltage. This is demonstrated in Figure 3b. In particular for $V_{bias} > 0.3$ V, the photocurrent $I_{photo}$ is proportional to the dark current. In other words, the conductance is increased when the laser excites the InAs-NW as shown for etched nanowires,[53,54] and that this photoconductance seems to dominate for a large bias voltage. For a bias voltage close to zero, the open circuit voltage of the Si/InAs heterojunction lets the photocurrent deviate from the dark current-voltage characteristic.

Figure 4 visualizes photocurrent maps for various bias voltages. The horizontal dashed white lines indicate the edge of the metal top contact. For the reverse direction ($V_{bias} < 0$ V, Figures 4a and 4e), the phase of $I_{photo}$ is constant and negative, *i.e.* $I_{photo}$ is negative for all positions. At a bias voltage $V_{oc,NW} < V_{bias} < V_{oc,Si}$, the photocurrent at the NW position already changes sign (Figures 4b and 4f). Figure 4 nicely demonstrates that it is crucial to record high-resolution photocurrent maps in order to distinguish the optoelectronic response of the Si/InAs heterojunction itself from that of the Si substrate, which is only drained through the InAs nanowire.

The differing optoelectronic response can also be seen in a distinct frequency dependence. Figure 5 shows the photocurrent at the NW and at the position "ref2" for three different chopper frequencies. We observe a reduction of the signal if we excite the silicon substrate ("ref2"), while the photocurrent characteristic at the NW is almost unaffected by the variation of the frequency. The frequency dependence of the signal at the position "ref2" shows that the corresponding charge carrier dynamics underlie a different generation mechanism than at the nanowire. In particular, the excitation energy of 0.84 eV ($\lambda = 1470$ nm) is below the bandgap of the Si substrate. Silicon is known to exhibits surface states $D_{Si}$ with a characteristic trapping time in the order of ms.[51] These traps exist at the surface of a silicon crystal.[36,50] Hereby, we interpret the photocurrent at the position "ref2" to stem from the excitation of the surface states $D_{Si}$ of the Si substrate. At the position of the NW, the optoelectronic dynamics are faster, which is consistent with recent time-resolved photocurrent experiments on single InAs NWs in THz stripline circuits.[55]

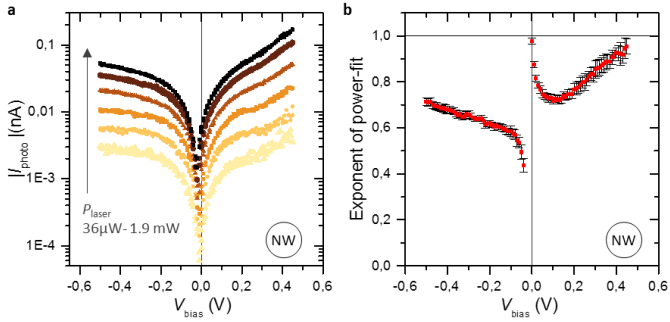

Figure 6 **Laser power dependence of the photocurrent:** **a.** Photocurrent $I_{photo}$ at the position of an InAs-NW *vs.* bias voltage for increasing laser power ($f_{chopper} \approx 833$ Hz, $T = 298$ K). For analysis, we fit the data with a simple power law $I_{photo} = A \cdot P_{laser}^{\beta}$. **B.** Exponent $\beta$ as a function of the bias voltage.

Figure 6a depicts the photocurrent at the InAs-NW position for different laser powers. We find that the power dependence varies for different $V_{bias}$. To further analyze the power dependence, we apply a simple fit function $I_{photo} = A \cdot P_{laser}^{\beta}$ and extract the exponent $\beta$ as a function of $V_{bias}$ (Figure 6b). This fit function describes most of the laser power dependence well. Deviations arise for $V_{bias} \approx V_{oc,NW} < 0$ V, where the photovoltaic contribution is dominating $I_{photo}$. The underlying optoelectronic processes will be discussed in the following section.

**Interpretation**

In electrical characterization measurements without laser excitation, Bessire *et al.* showed that mid-gap states $D_{InAs}$ in the InAs can generate an additional transport path at a Si/InAs heterojunction.[30] In the following, we consider the impact of such mid-gap trap states $D_{InAs}$ at the Si/InAs heterojunction on the photocurrent generation. From the negative open circuit voltage at the NW position, we conclude that the slope of the band bending in the InAs must be opposite to the slope of the band bending in the silicon, *i.e.* the built-in electric field in the silicon as well as in the InAs point away from the heterojunction. For an ideal Si/InAs heterojunction without interfacial traps and $V_{bias} = 0$ V, the photogenerated electrons in the InAs-NW can only contribute to the photocurrent if they move into the silicon substrate. However, the barrier $\Delta E_C$ is too high for thermalized electrons to overcome. We therefore infer that photogenerated electrons recombine with holes from the p-type silicon *via* the trap states $D_{InAs}$ in the InAs, as illustrated by the orange arrows in Figure 2d.

We note that we find a negative open circuit voltage only when we excite the Si/InAs heterojunctions below the band gap of silicon and with a passivated silicon surface. With an excitation above the silicon band gap, we observe a positive open circuit voltage (data not shown) in agreement with recent observations in literature.[56] Moreover, we find that the photovoltaic generated current at $V_{bias} = 0$ V increases linearly with laser power (Figure 6b). This demonstrates that the number of traps states does not limit the drain of the photocurrent. This conclusion is supported by another argument. The current voltage characteristics of Figure 2a can be fitted with a simple parallel circuit (inset of Figure 2a), consisting of a diode and a shunt resistance $R$. The reverse current ($V_{bias} < 0$ V) is well described by this resistance. Figure 2a shows that the ohmic contribution in reverse direction decreases with temperature. We find that the temperature dependence of the shunt resistance $R(T)$ follows a Boltzmann distribution (Figure in supporting material Arrhenius $R(T)$), *i.e.* here, the charge carriers are thermally excited from the valence band to the conduction band. In forward direction, the electrons in the InAs again need to recombine with holes of the silicon *via* the trap states $D_{InAs}$. Obviously the trap states $D_{InAs}$ do not hinder the relative high ohmic forward current flow. This conclusion is consistent with several studies showing that the trap state density can be high enough to allow such current flow across the heterojunction.[30–32]

Figure 3b demonstrates that both $I_{photo}$ and $I_{dark}$ resemble each other for $V_{bias} > 0.3$ V. In addition, the increase is exponential, as expected for a diode. For higher forward biases, we therefore infer that the photocurrent stems from an increased photoconductance, which is in accordance with the approximately linear power dependence of $I_{photo}$ for $V_{bias} > 0.3$ V: The Fermi level of InAs is pinned at its surface. Because of this, electrons will accumulate on the surface, whereas holes tend to move into the InAs NW core (inset of Figure 2b). Upon optical excitation, the number of charge carriers is increased in the InAs-NW, and the accumulation of electrons at the InAs-NW surface increases the hole channel diameter within the NW core. Thereby, an optically induced gating of the NW is created that increases the conductance (*i.e.* the cross section). In turn, the bias dependence of the photocurrent and the dark current resemble each other.[54,57,58]

In reverse direction, photogenerated electrons can drain from the InAs NW to the top contact, and holes only need to overcome a small barrier in the InAs NW. Therefore, one would expect a relatively high photocurrent under the reverse biasing. However, the energetic barrier for holes $\phi_{rev}$ inside the InAs is increased because of the surface Fermi level pinning at the InAs NW (compare Figure 2d and inset of Figure 2b). The dashed lines on the right in Figure 2d schematically depict the energetic lifting of the band edges in the InAs NW core. This potential barrier $\phi_{rev}$ hinders holes to drain into the silicon under reverse biasing.

The laser power dependence corroborates the above interpretation. For $V_{bias} = 0$ V (Fig. 6b), we find that the power exponent is close to one ($\beta \approx 1$), *i.e.* the photocurrent close to the short circuit condition increases linearly with the laser power. This observation is consistent with a photovoltaic effect with a negative open circuit voltage measured at the position of the NW. In other words, the photocurrent at zero bias stems from photogenerated charge carriers that are dissociated by electric fields at the Si/InAs heterojunction.[10–13] For biases $0.1 \leq |V_{bias}| \leq 0.4$ V, the power dependence is sublinear ($\beta < 1$). This indicates that the photocurrent is no longer solely due to a photovoltaic effect, but that it is rather caused by trap state dynamics within the InAs-NW.[59–61] For large bias voltages $V_{bias} > 0.4$ V, the exponent $\beta$ increases towards 1. The trend is consistent with a dominating photoconductance effect at large biases, which is in accordance with earlier results on III/V NWs.[57]

# Conclusion

We observe that a transport path through mid-gap defects can dominate the photocurrent dynamics across single Si/InAs heterojunctions. We demonstrate that the defects can arise from periodic misfit dislocations at Si/InAs heterojunction in the InAs-NW. Furthermore, we find a relative small and negative open circuit voltage of −21 mV, which shows that a photovoltaic effect across Si/InAs heterojunctions only dominates for small voltages. From the negative sign of the open circuit voltage, we can conclude that interfacial charges determine the sign of the band bending at the heterojunction. For larger bias voltages, we infer that the optoelectronic response of the junction stems from a photogating effect of the InAs-NW that increases the conductivity of the InAs-NW. Finally, we address the role of the defects at the silicon substrate interface on the optoelectronic response. Since III/V-NWs on silicon platforms are considered for future photovoltaic and optoelectronic circuits, our results may prove essential for the understanding and the design of such devices based on NW-silicon heterojunctions.

# Methods

### Sample preparation

A 20-nm thick thermal $SiO_2$ on top of (111) oriented (of ± 0.5 deg uncertainty), p-type Si (specific resistance 1 - 30 Ohm cm) was patterned with electron beam lithography and subsequent wet chemical etching (10% HF solution) to form circular openings. After transfer to the MBE growth chamber, free-standing InAs NWs were nucleated and grown from the prepatterned openings using growth parameters of $T$ = 500 °C, In-rate 0.26 Å/s, As-flux 3.5 × $10^{-5}$ mbar, and growth duration of 10 min. The NWs grew along the [111] direction with a very slight tilt of ≈ 0.6 deg. The NWs had a typical length and diameter of (1.2 ± 0.1) um and (45 ±5) nm, respectively. The latter value was determined by averaging SEM images of several NWs. Note, that in particular for thin NWs, the SEM contrast is limited and slightly overestimates the NW thickness. After MBE-growth, the sample was again etched in buffered HF solution and subsequently spincoated with a transparent insulating polymer (Bisbenzocyclobutene, BCB) to completely encapsulate the NW. The BCB was thermally cured at 240°C for 6 minutes and subsequently etched with a $C_4F_8/SF_6$ plasma in a reactive ion etching system (RIE, Oxford Instruments Plasmalab 80 Plus) to expose the InAs NW tips. The exposure of the tips was verified using AFM measurements. Finally, a thin, semitransparent contact was evaporated on top (5 nm titanium and 25 nm gold) to cover the complete NW tip. The sample was glued with silver paste into a chip carrier and wire bonded for electrical read-out.

## Optoelectronic set-up

The sample was mounted in to a vacuum chamber and was excited with a continuous-wave-laser beam (λ = 1470 nm), that was focused with a ×50 objective. The position of the focused laser spot on the sample (diameter ≈ 5 µm) was scanned with a three-axes piezo stack. The laser irradiation was modulated with a chopper wheel in the kHz-frequency range. A lock-in amplifier was used to measure the photocurrent according to the chopper modulation. The current signal was pre-amplified with an Ithaco 1211 current voltage converter.

## Corresponding Author

holleitner@wsi.tum.de

## Acknowledgements

This work was supported by a European Research Council (ERC) Grant NanoREAL (No. 306754), and the IGGSE project "Advanced III/V semiconductor nanowire-based electronics".

# Supporting Information

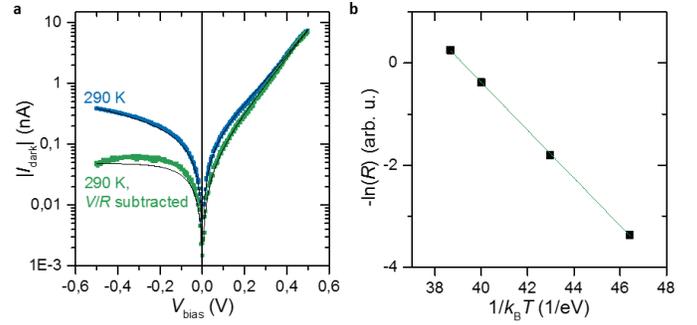

Figure S 1 **Temperature Dependence of the ohmic contribution of the dark current:** We fit the temperature dependent data of Figure 2a of the main manuscript with the equivalent circuit shown in the inset of Figure 2a: $I = I_1 \cdot \left(\exp\left\{e \cdot \frac{V_{\text{bias}}}{\eta k_\text{B} T}\right\} - 1\right) + \frac{V_{\text{bias}}}{R}$. Supporting Figure **a** shows the fit (solid line) as well as the data (blue data points) and the fit after the subtraction of the ohmic contribution $\frac{V_{\text{bias}}}{R}$ (solid line and green data points, respectively). The shunt current through the trap states depends on the thermally distributed charge carrier density $n \propto \exp\left\{-\frac{E_\text{g}}{2k_\text{B}T}\right\} \propto 1/R$. The supporting figure **b** shows the dependence of $R(T)$ in an Arrhenius plot. We find a bandgap energy of $E_\text{g} \approx 0.94$ eV $> E_\text{g,InAs}$, which refers to the thermal activation of charge carriers in silicon ($E_\text{g,Si} \approx 1.1$ eV) and the InAs ($E_\text{g,InAs} \approx 0.36$ eV).